\documentclass[sigconf,nonacm,natbib=true]{acmart}
\AtBeginDocument{%
  }

\usepackage{csquotes}
\usepackage{mdframed}
\usepackage{enumitem}
\usepackage{tabularx}
\usepackage{xcolor}
\usepackage{booktabs}
\usepackage{url}

\newcommand{\systemname}{\textsc{SimEval-IR}}

\newmdenv[
  backgroundcolor=gray!8,
  linecolor=gray!50,
  roundcorner=4pt,
  skipabove=6pt,
  skipbelow=6pt,
  innerleftmargin=8pt,
  innerrightmargin=8pt,
  innertopmargin=6pt,
  innerbottommargin=6pt
]{prompt}


\begin{document}

\title[\textsc{SimEval-IR}: A Unified Toolkit and Benchmark Suite for Evaluating User Simulators and Search Sessions]{\textsc{SimEval-IR}: A Unified Toolkit and Benchmark Suite for Evaluating User Simulators and Search Sessions}

\author{Saber Zerhoudi}
\orcid{0000-0003-2259-0462}
\affiliation{%
  \institution{University of Passau}
  \city{Passau}
  \country{Germany}
}
\email{szerhoudi@acm.org}

\makeatletter
\setlength{\skip\footins}{9pt plus 2pt minus 1pt}

\newcommand{\acmrightssize}{\fontsize{8}{9.5}\selectfont}

\setlength{\emergencystretch}{1.5em} 

\settopmatter{printacmref=false}

\newcommand{\firstpagerights}[1]{%
  \begingroup
    \renewcommand\thefootnote{}%
    \footnotetext{%
      \acmrightssize
      \raggedright
      \setlength{\parskip}{0pt}%
      \setlength{\parindent}{0pt}%
      #1%
    }%
    \addtocounter{footnote}{0}%
  \endgroup
}
\makeatother

\begin{abstract}
User simulators are increasingly central to interactive information retrieval, yet the community lacks standardized evaluation tools. Simulators serve two objectives, behavioral realism (matching real user behavior) and tester reliability (producing valid system rankings), and these are often conflated despite being distinct and sometimes conflicting. We present \systemname{}, an open-source toolkit and benchmark suite that makes this distinction measurable. \systemname{} provides: (1)~a canonical session schema unifying session search and conversational interactions, with validated dataset adapters and explicit loss accounting; (2)~three executable benchmarks covering behavioral realism, tester reliability with RATE-style estimation, and an analysis linking the two; and (3)~baseline results across four real datasets in two languages and four simulator families. Our key finding: the classifier-discriminator ``human-likeness'' check, the dominant realism test in the literature, has essentially no pooled predictive power for system-ranking validity ($r{=}{+}0.09$, $n{=}48$), while marginal click-depth distance and Fr\'{e}chet distance over session embeddings give a much stronger signal ($|r|{=}0.43$ and $0.40$, $p{\leq}0.005$). \systemname{} is released with all configurations and scripts to reproduce the reported analysis.
\end{abstract}



\maketitle
\enlargethispage{2\baselineskip}
\firstpagerights{%
  © ACM, 2026. This is the author's version of the work.\\
  The definitive version was published in:
  \emph{Proceedings of the 49th International ACM SIGIR Conference on Research and Development in Information Retrieval (SIGIR '26), July 20--24, 2026, Melbourne, VIC, Australia}.\\
  DOI: \url{https://doi.org/10.1145/3805712.3808635}
}

\section{Introduction}
\label{sec:intro}

\begin{figure}[t]
    \includegraphics[width=\linewidth]{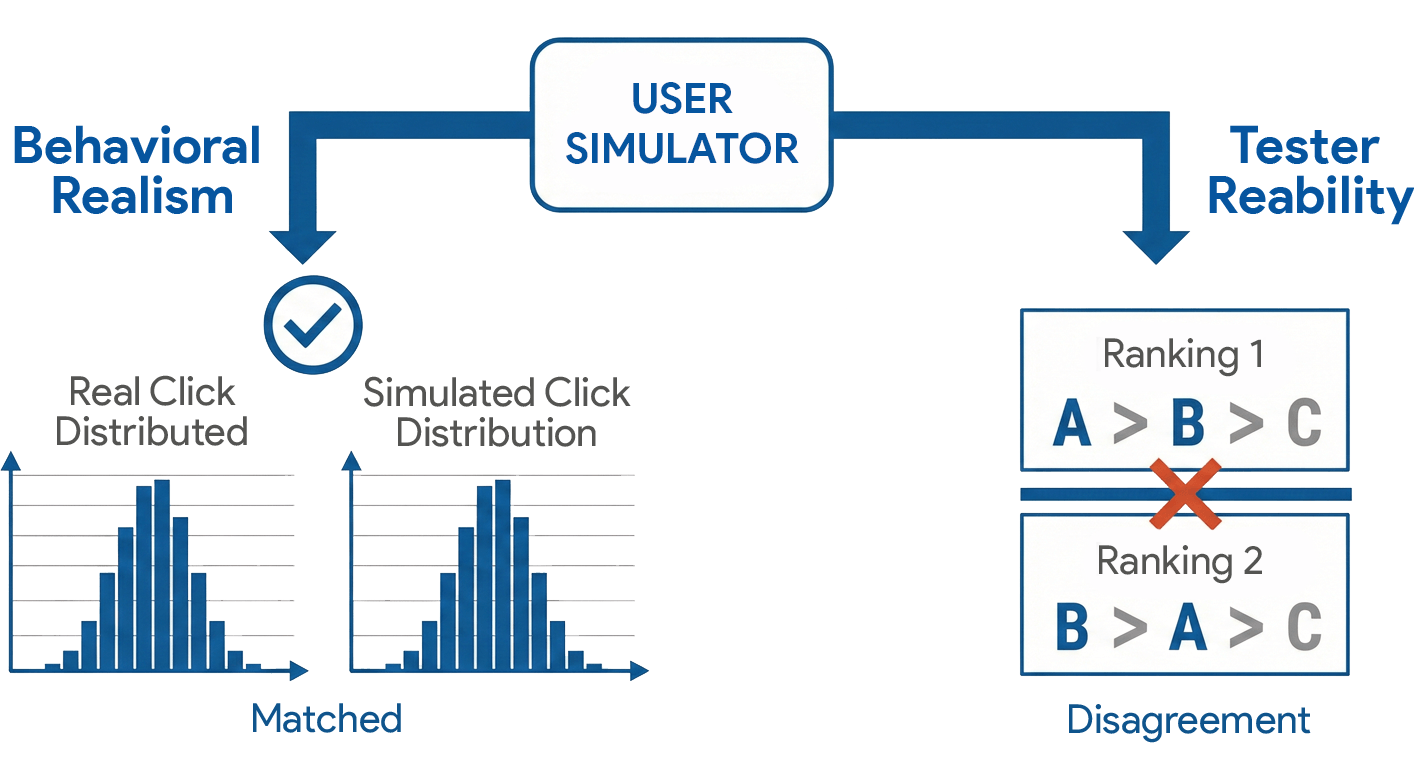}
    \caption{The same simulator can appear realistic under behavioral metrics while producing unreliable system rankings. \systemname{} standardises both evaluations.}
    \label{fig:motivation}
\end{figure}

Consider a researcher developing a new user simulator for interactive search. To evaluate it, they compute click-depth histograms, find a close match to real logs, and conclude the simulator is realistic \cite{Zerhoudi2022EvaluatingSimulated}. A second researcher, using a comparable simulator as a tester to compare retrieval systems, finds that system rankings diverge from those obtained with human assessors \cite{LabhishettyZhai2021TesterBased}. Both researchers followed standard practice in their own communities and reached different conclusions about the same class of simulator (Figure~\ref{fig:motivation}). This is not hypothetical: evaluation practices for behavioral realism \cite{Zerhoudi2022EvaluatingSimulated,Zerhoudi2024AssessingSimulators} and tester reliability \cite{LabhishettyZhai2021TesterBased,LabhishettyZhai2022RATE,BalogZhai2024UserSimulation} have grown up in parallel, with different datasets, metrics, and quiet assumptions about what makes a good simulator. The community has no shared infrastructure to tell these objectives apart or measure them jointly.

Recent work clarifies this distinction \cite{BalogZhai2024UserSimulation,LabhishettyZhai2021TesterBased}. A simulator serving as a \emph{behavioral generator} should match real user behavior in distributional and sequential properties. A simulator serving as a \emph{tester} should produce system rankings that agree with trusted evaluators. These objectives can conflict: a simulator may look plausible and still rank systems badly, or it may differ in surface statistics yet still rank systems correctly. Current evaluation defaults to ad hoc checks (matching session-length distributions, computing classifier-based ``Turing tests'' \cite{Zerhoudi2022EvaluatingSimulated}, or reporting occasional rank correlations \cite{LabhishettyZhai2021TesterBased}) that vary across papers in preprocessing, sessionization, and metric implementation. A high classifier AUC can also reflect pipeline quirks (missing fields, timestamp conventions), not behavior.

We present \textbf{\systemname{}}, a simulator-centric evaluation toolkit and benchmark suite that makes this distinction measurable. \systemname{} contributes:

\begin{itemize}[label=\textendash,labelindent=0pt,leftmargin=2em,labelsep=0.4em]
    \item A \textbf{canonical session schema} representing query--SERP--click sessions and conversational interactions, with validated dataset adapters and explicit loss accounting that records what is preserved, dropped, or missing during preprocessing.
    \item Three \textbf{executable benchmark tasks}: (B1)~behavioral realism, (B2)~tester reliability with RATE-style estimation \cite{LabhishettyZhai2022RATE}, and (B3)~analysis of when and why realism predicts reliability.
    \item \textbf{Reproducible baselines} computed end-to-end on four real datasets across two languages: TREC Session Track 2014 (1{,}257 real graded sessions \cite{Carterette2016TrecSession}), the full AOL-IA query log (36.4M qlog entries, 19.4M qrels \cite{AOLQueryLog2006}), the MS MARCO ORCAS click log (10.4M queries, 18.8M qrels \cite{Craswell2020ORCAS}), and Chinese MIRACL/zh \cite{Zhang2023MIRACL}, using literature-parameterised PBM \cite{Craswell2008PBM} and DBN \cite{ChapelleZhang2009DBN} click models, a position-only heuristic, and a stylised LLM-style simulator. The B3 analysis aggregates 48 simulator-seed configurations across two retrieval-grounded session corpora.
    \item An \textbf{open-source release} with configurations, adapters, and scripts that reproduce every analysis from the raw qlog.\footnote{\url{https://github.com/searchsim-org/sigir26-simeval-ir}}
\end{itemize}

\systemname{} is a reference benchmark that standardises \emph{what} to evaluate, \emph{how} to evaluate it, and \emph{how} to report results. Validity rules are checked when a benchmark runs: metrics that need qrels cannot run on datasets without judgments, classifier AUC is always reported next to its baselines so the score can be read in context, and every run writes a configuration hash and a provenance file.

\section{Background and Positioning}
\label{sec:background}

\systemname{} builds on three lines of prior work: interaction datasets, evaluation toolkits, and simulator-centric evaluation theory. We review each with emphasis on what resources do not standardize.

\textbf{Interaction datasets.}
Web query logs (AOL \cite{AOLQueryLog2006}, Yandex, Sogou) provide large-scale behavioural signals but differ in schemas and sessionisation conventions. Domain-specific logs (TripClick \cite{Rekabsaz2021TripClick}, TianGong-ST \cite{Chen2019TianGong}) add diversity but bring their own quirks. TREC-style resources add reproducible evaluation: the Session Track \cite{Carterette2016TrecSession} and CAsT \cite{Dalton2020TrecCast} provide qrels. Dialogue corpora and LLM chat logs \cite{Lin2025WildBench,LMSYSChat1M} are increasingly used to build LLM-based simulators, yet many lack retrieval grounding (no SERPs, no qrels), so metrics meant for one setting end up being applied to another. \systemname{} treats these resources as \emph{distinct evaluation regimes} and records which metrics are valid in each.

\textbf{Evaluation toolkits.}
Toolkits such as \texttt{\small trec\_eval} \cite{trec_eval}, \texttt{\small ir-measures} \cite{MacAvaney2022IrMeasures}, and PyTerrier \cite{Macdonald2021PyTerrier} provide implementations for ranked-list metrics but target query/document/qrels pipelines, not session logs and simulator outputs. Researchers studying interaction resort to custom preprocessing and one-off scripts.

\textbf{Simulator-centric evaluation.}
Behavioural realism is typically measured by distributional comparisons or classifier-based ``discriminator'' tests \cite{Zerhoudi2022EvaluatingSimulated,Zerhoudi2024AssessingSimulators}; tester-based evaluation asks whether the rankings a simulator produces agree with a trusted evaluator \cite{LabhishettyZhai2021TesterBased}. RATE \cite{LabhishettyZhai2022RATE} formalises reliability estimation and aggregation, and recent commentary \cite{BalogZhai2024UserSimulation} points out that behavioural plausibility does not imply ranking validity. Despite this clarity, no reference benchmark covers both objectives across multiple datasets. A common hazard is classifier-based realism: a classifier can pick up pipeline quirks (missing dwell fields, timestamp quantisation, dataset-specific metadata) and not behavior, so any classifier-AUC score should be reported next to the artifact-only baselines that show what part of the score is real.

\section{\systemname{} Overview}
\label{sec:overview}

\begin{figure*}[t]
    \centering
    \begin{minipage}[b]{0.60\textwidth}
        \centering
        \includegraphics[width=\linewidth]{figures/simeval_06.png}
        \caption{\systemname{} workflow. Adapters convert heterogeneous inputs into a canonical schema with explicit loss accounting. Three benchmarks evaluate behavioral realism (B1), tester reliability (B2), and their relationship (B3).}
        \label{fig:architecture}
    \end{minipage}\hfill
    \begin{minipage}[b]{0.38\textwidth}
        \centering
        \includegraphics[width=\linewidth]{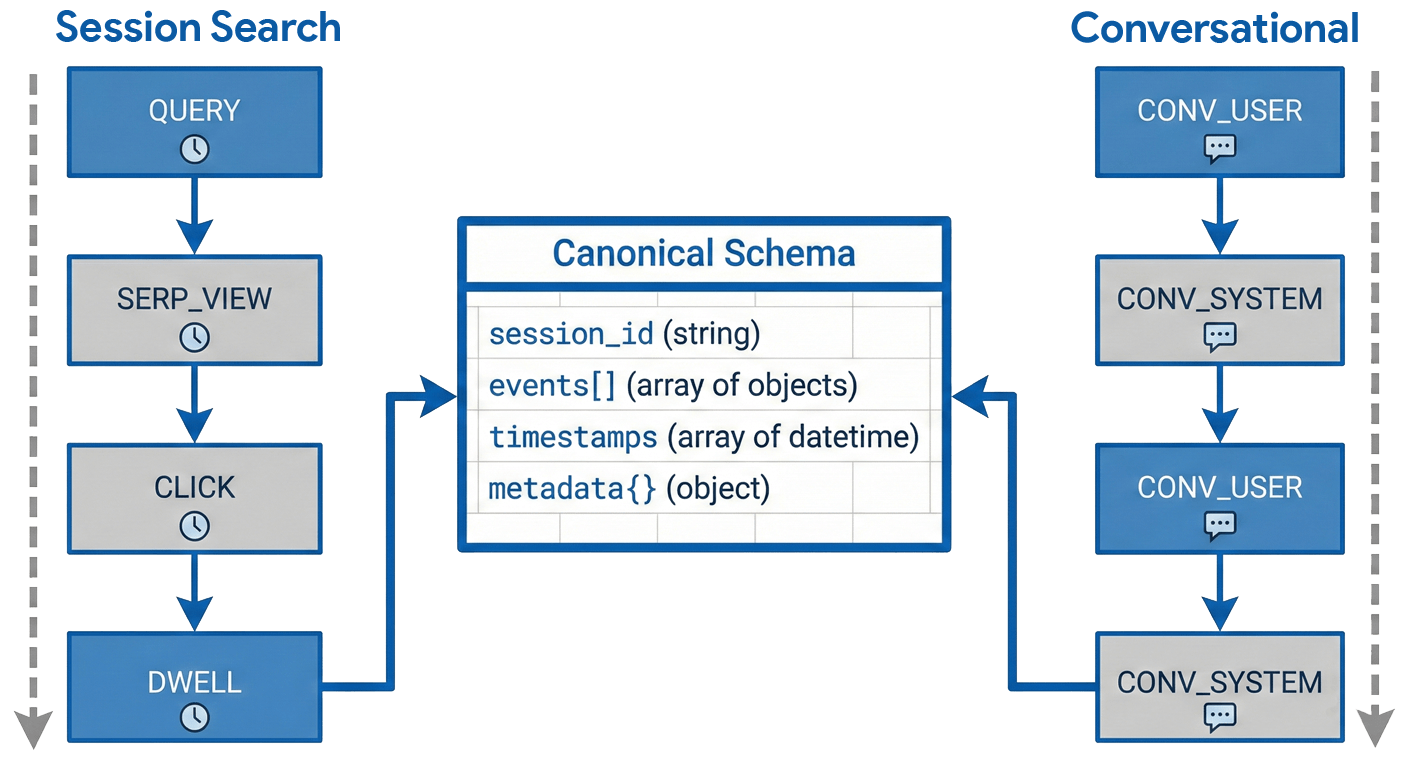}
        \caption{Canonical session schema unifying session search and conversational interactions as ordered event sequences with typed events.}
        \label{fig:schema}
    \end{minipage}
\end{figure*}

\systemname{} standardises how sessions are represented, how datasets and simulator outputs are converted into that representation, and how simulators are evaluated under distinct objectives. Figure~\ref{fig:architecture} illustrates the workflow.

\systemname{} targets two interaction settings with different evaluation needs. \emph{Retrieval-grounded} datasets (session search, conversational search with judged turns) support realism and reliability analysis (B1, B2, B3). \emph{Dialogue-only} corpora lack SERPs and judgements, so only B1 (structural realism) applies. The framework's job is to stop invalid comparisons: B2 and B3 will not run on datasets without qrels, and dialogue-only corpora are excluded by default.

\textbf{Canonical session schema.}
Sessions are ordered events typed as \texttt{QUERY}, \texttt{SERP\_VIEW}, \texttt{CLICK}, \texttt{DWELL}, \texttt{CONV\_USER}, or \texttt{CONV\_SYSTEM} (Figure~\ref{fig:schema}). The schema is versioned and validated; optional fields let datasets that lack certain signals fit without forced conversion.

\textbf{Dataset adapters with loss accounting.}
Each adapter converts a dataset's native format into the canonical schema and writes a manifest that records session and event counts, missingness rates, dropped fields, and configuration hashes. This makes preprocessing auditable and supports leakage checks.

\textbf{Metric library.}
Metrics are grouped into three families: (M1)~behavioural realism (distributional distances, sequential metrics, embedding-based distances, classifier discrimination), (M2)~session effectiveness (when qrels exist), and (M3)~tester reliability (ranking agreement, RATE-style estimation \cite{LabhishettyZhai2022RATE}).

\textbf{Executable benchmarks.}
Three tasks produce structured reports (JSON, tables, figures) designed for direct cross-paper comparison. Each is defined by a YAML configuration specifying datasets, preprocessing, features, and metrics.

\paragraph{\textbf{B1: Behavioral Realism.}}
Given real and simulated sessions under the same schema, B1 computes distributional, sequential, and representation-level realism metrics with confidence intervals, alongside classifier-based discrimination. Because a classifier can pick up dataset quirks, the classifier's main AUC is reported next to four reference baselines: (i)~a metadata-only baseline trained on schema-presence indicators only, (ii)~a structural-only model trained on action-sequence statistics, (iii)~a masked-feature ablation that removes session-length features, and (iv)~a permutation check with shuffled labels. The main AUC reflects a real behavioural difference only when (i) and (iv) stay near chance.

\paragraph{\textbf{B2: Tester Reliability.}}
Given a set of system runs, B2 evaluates each system under each tester (qrels-based, log-based, simulator-based) and computes ranking agreement (Kendall's~$\tau$, Spearman's~$\rho$, Pearson) with bootstrap confidence intervals. RATE-style estimation \cite{LabhishettyZhai2022RATE} iteratively re-weights testers by their correlation with the consensus ranking, downweighting noisy testers and producing an aggregated ranking; we converge in $\leq{}3$ iterations on our setup. Leave-one-out sensitivity flags rankings that depend on a single tester being in the panel.

\paragraph{\textbf{B3: Realism--Reliability Analysis.}}
B3 links B1 and B2 by measuring how well realism metrics predict tester reliability across simulators and seeds. This benchmark targets the central question \systemname{} is designed to answer: \emph{which realism metrics, if any, are informative predictors of evaluation validity?}

\section{Dataset Coverage and Validation}
\label{sec:datasets}

\systemname{} ships ten openly-accessible dataset adapters (Table~\ref{tab:datasets}), plus adapters for TripClick \cite{Rekabsaz2021TripClick} and TianGong-ST \cite{Chen2019TianGong} where users supply the source files. The baseline experiments in Section~\ref{sec:results} run end-to-end on four corpora: TREC Session Track 2014 \cite{Carterette2016TrecSession} (1{,}257 sessions, 16{,}949 graded qrels), the full AOL-IA release \cite{AOLQueryLog2006} (36.4M qlog entries, 19.4M binary qrels), the MS MARCO ORCAS click log \cite{Craswell2020ORCAS} (10.4M queries, 18.8M qrels), and the Chinese MIRACL/zh dev split \cite{Zhang2023MIRACL} (393 queries, 3{,}928 qrels).

\begin{table}[t]
\centering
\small
\caption{Dataset coverage. ``G''=retrieval grounding; ``J''=judgements; ``S''=full session structure. B2/B3 require ``J''; B1 with real-vs-sim comparison requires ``S''. The four bold rows are the testbeds used in Section~\ref{sec:results}.}
\label{tab:datasets}
\begin{tabular*}{\columnwidth}{@{\extracolsep{\fill}}lccccc@{}}
\toprule
Dataset & Type & G/J/S & Lang & B1 & B2 \\
\midrule
\textbf{TREC Session'14} \cite{Carterette2016TrecSession} & Session & \checkmark / \checkmark / \checkmark & EN & \checkmark & \checkmark \\
\textbf{AOL-IA} \cite{AOLQueryLog2006} & Session & \checkmark / \checkmark / \checkmark & EN & \checkmark & \checkmark \\
\textbf{ORCAS} \cite{Craswell2020ORCAS} & Click log & \checkmark / \checkmark / -- & EN & -- & \checkmark \\
\textbf{MIRACL/zh} \cite{Zhang2023MIRACL} & Passage & \checkmark / \checkmark / -- & ZH & -- & \checkmark \\
\midrule
Yandex & Session & \checkmark / -- / \checkmark & RU & \checkmark & -- \\
Sogou & Session & \checkmark / -- / \checkmark & ZH & \checkmark & -- \\
TREC CAsT \cite{Dalton2020TrecCast} & Conv. & \checkmark / \checkmark / \checkmark & EN & \checkmark & \checkmark \\
Persona-Chat & Dialogue & -- / -- / \checkmark & EN & \checkmark & -- \\
LMSYS-Chat-1M \cite{LMSYSChat1M} & Dialogue & -- / -- / \checkmark & Multi & \checkmark & -- \\
WildBench \cite{Lin2025WildBench} & Dialogue & -- / -- / \checkmark & Multi & \checkmark & -- \\
\bottomrule
\end{tabular*}
\vspace{-0.6em}
\end{table}

For datasets without explicit session boundaries, \systemname{} applies configurable inactivity-timeout sessionisation (30~minutes by default for web logs, 60~minutes for academic search), with both choices recorded in the manifest. Adapters validate materialised sessions against schema constraints: click ranks must fall within SERP bounds, timestamps must be non-decreasing, and item identifiers must be consistent. As an example of the value of loss accounting, ingesting AOL-IA drops 14{,}311 single-query traces below the minimum-session-length threshold, a figure rarely reported in papers using this dataset but that directly affects distributional realism comparisons.

\textbf{Simulator integration.}
\systemname{} provides a simulator adapter interface compatible with SimIIR \cite{Zerhoudi2022SimIIR2,azzopardi2024simiir}; we verified interoperability by ingesting the Sim4IA-Bench Task-A release \cite{kruff2026sim4ia} into the canonical schema and running the B1 pipeline against it. For LLM-based simulators, the interface requires event traces consistent with the schema, treating prompt and action space as recorded configuration artifacts; the release also ships a reference real-LLM driver that runs against any OpenAI-compatible endpoint.

\section{Metrics and Benchmark Protocols}
\label{sec:metrics}

\systemname{} defines metrics with applicability constraints enforced at runtime.

\textbf{Behavioral realism metrics.}
\emph{Marginal distribution metrics} compare single-feature distributions (session length, click depth, dwell time, query length) via Jensen--Shannon divergence, Wasserstein distance, and Kolmogorov--Smirnov statistics. \emph{Sequential metrics} capture transition dynamics: Jensen--Shannon divergence over action-bigram transition matrices, normalised Levenshtein distance between action-type sequences, and reformulation similarity. \emph{Representation-level metrics} compute Maximum Mean Discrepancy (MMD) \cite{Gretton2012MMD} and Fr\'{e}chet distance (FD) \cite{Zerhoudi2024AssessingSimulators}, originally proposed for comparing feature-space distributions of generative models \cite{Heusel2017FID}, on learned session embeddings. The default embedder is a 156-dimensional action-sequence representation (event-type frequencies plus transition probabilities); the protocol accepts arbitrary embedders \cite{Reimers2019SBERT}.

\begin{figure}[t]
    \centering
    \includegraphics[width=.78\linewidth]{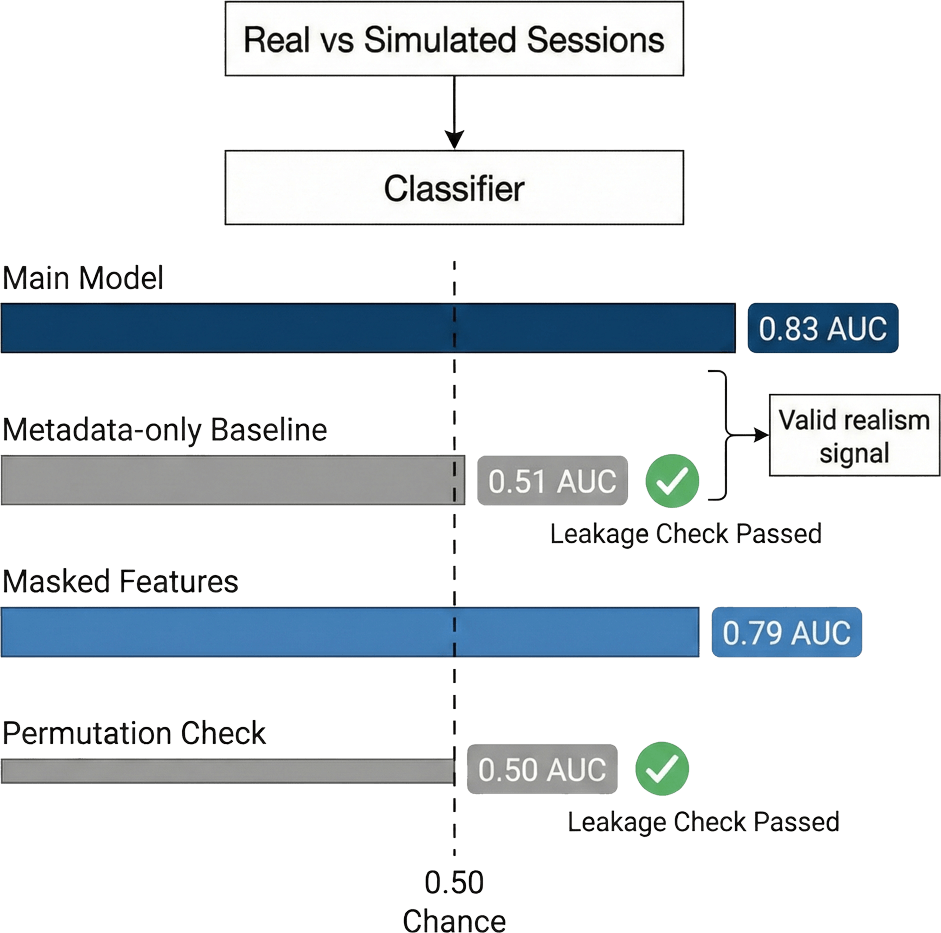}
    \caption{Classifier-discriminator interpretation: the main AUC is meaningful only when the metadata-only and permutation baselines stay near chance.}
    \label{fig:leakage}
    \vspace{-1.6em}
\end{figure}

\textbf{Classifier realism with mandatory leakage audits.}
A binary classifier trained to tell real and simulated sessions apart can reach high accuracy by picking up pipeline quirks (timestamp conventions, missing-field signatures) rather than behaviour. The four reference baselines below are therefore part of the metric itself and are reported alongside the main AUC (Figure~\ref{fig:leakage}):

\begin{enumerate}[leftmargin=*,itemsep=1pt,topsep=2pt]
    \item \emph{Metadata-only}: trained on schema-presence indicators only. High AUC means the classifier is learning the adapter signature, not behaviour.
    \item \emph{Structural-only}: action-sequence statistics in isolation, to isolate the contribution of session structure.
    \item \emph{Masked-feature}: removes session-length features so length effects cannot dominate the score.
    \item \emph{Permutation}: random labels; AUC must collapse to chance.
\end{enumerate}

\noindent The main AUC is read as a behavioural difference only when (1) and (4) stay near chance ($\text{AUC}{<}0.55$); otherwise the realism score is dropped from the report.

\textbf{Tester reliability metrics.}
We compute Kendall's~$\tau$, Spearman's~$\rho$, Pearson correlation, top-weighted $\tau_{AP}$ \cite{Yilmaz2008TauAP}, and pairwise system concordance against a trusted evaluator (qrels-nDCG@10), each with bootstrap 95\% confidence intervals. RATE \cite{LabhishettyZhai2022RATE} estimates per-tester reliability by alternating between (i)~a weighted consensus ranking and (ii)~each tester's correlation with the consensus, until weights converge. The aggregated ranking downweights noisy testers; leave-one-out sensitivity flags fragile rankings where removing one simulator changes the top-ranked system.

\section{Baseline Results and Key Findings}
\label{sec:results}

All numbers in this section are produced by the released scripts and configuration files; bit-for-bit reproduction requires only the AOL-IA release and seeds 20260429--20260504.

\subsection{Experimental Setup}

\textbf{B1 reference corpora} are the full TREC Session Track 2014 release (1{,}257 real sessions, 3{,}645 interactions, 1{,}041 click events) and AOL-IA, sliced into six disjoint 2{,}000-session shards so per-seed B1 estimates are independent. Four simulators run on the same SERPs as the real sessions and only re-sample the click events. \emph{SimIIR-PBM} uses literature-standard position-based parameters \cite{Craswell2008PBM} with rank-attractiveness $\{0.05,0.55\}$; \emph{SimIIR-DBN} \cite{ChapelleZhang2009DBN} adds cascade satisfaction with $\gamma{=}0.20$; \emph{Heuristic} uses pure rank-based clicks $p_r{\propto}1/r$ that ignore relevance; \emph{LLM-sim} is a stylised LLM-style click model that examines deeper ranks, inflates dwell times, and verbose-paraphrases follow-up queries. Exact hyperparameters and simulator code are in the released configuration files.

\textbf{B1 metrics.} 5-fold cross-validated logistic regression on action-sequence embeddings; FD and MMD are computed on the same 156-dimensional embeddings.

\textbf{B2 testbeds.} For each dataset we sample 100 queries (40 topics for TREC S14) with their qrels, build a candidate pool padded with distractor docs where the qrels are positive-only (AOL, ORCAS, MIRACL/zh), and synthesise 10 retrieval systems by ranking pools with quality $\alpha\in[0.15, 0.80]$ (an $\alpha$-weighted mixture of true relevance and Gaussian noise). Each tester replays its click model 8 times per (system, query) and reports mean click-derived nDCG@10. The trusted tester is qrels-nDCG@10 on the same pools.

\subsection{Behavioral Realism (B1)}

Table~\ref{tab:b1} reports B1 on the canonical seed for both retrieval-grounded session corpora. The metadata-only baseline stays at AUC $\leq 0.50$ and the permutation baseline at AUC $\leq 0.55$ for all eight configurations, so the main classifier AUC values reflect real behavioural discrepancy and not schema artifacts.

\begin{table}[t]
\centering
\small
\caption{B1 realism on TREC Session'14 (top, 1{,}257 real sessions, graded qrels) and AOL-IA (bottom, 2{,}000 real sessions, binary qrels). Lower is better for distances; higher AUC = worse realism. Metadata-only and permutation baselines are below chance for all configurations. }

\label{tab:b1}
\begin{tabular*}{\columnwidth}{@{\extracolsep{\fill}}lcccccc@{}}
\toprule
Simulator & JS\,(click) & W\,(len) & Fr\'{e}chet & MMD & Reform & AUC \\
\midrule
\multicolumn{7}{c}{\textit{TREC Session Track 2014}} \\
SimIIR-PBM & 0.117 & 0.50 & 0.048 & 0.158 & 0.000 & 0.577 \\
SimIIR-DBN & 0.203 & 1.58 & 0.389 & 0.447 & 0.000 & 0.768 \\
Heuristic  & 0.154 & 1.39 & 0.331 & 0.417 & 0.000 & 0.747 \\
LLM-sim    & 0.211 & 1.70 & 0.518 & 0.505 & 0.100 & 0.804 \\
\midrule
\multicolumn{7}{c}{\textit{AOL-IA}} \\
SimIIR-PBM & 0.197 & 1.80 & 0.577 & 0.430 & 0.000 & 0.910 \\
SimIIR-DBN & 0.013 & 1.02 & 0.217 & 0.258 & 0.000 & 0.822 \\
Heuristic  & 0.304 & 2.11 & 0.823 & 0.513 & 0.000 & 0.919 \\
LLM-sim    & 0.090 & 1.60 & 0.470 & 0.391 & 0.291 & 0.896 \\
\bottomrule
\end{tabular*}
\vspace{-0.6em}
\end{table}

The two corpora tell different stories. On TREC Session'14, SimIIR-PBM is the closest match on every distance metric (JS, W, FD, MMD) and produces the lowest classifier AUC (0.577), the only configuration close to chance; the graded qrels and short sessions match the PBM independence assumption. On AOL-IA, SimIIR-DBN is the closest match on every distance metric and produces a lower classifier AUC than LLM-sim (0.822 vs.\ 0.896), even though 2{,}000 long, multi-query sessions give the classifier more to work with; the cascade-satisfaction pattern DBN models is what AOL sessions look like. No simulator wins on both datasets, which is why a multi-corpus profile beats a single ``realism score''.

\subsection{Tester Reliability (B2)}

Table~\ref{tab:b2} shows ranking agreement against the qrels evaluator on 10 retrieval systems for all four datasets. Across every corpus, every relevance-aware simulator (PBM/DBN/LLM) achieves \textbf{$\tau\geq{}0.77$ in 12/12 cases} while the position-only Heuristic baseline collapses to chance everywhere except the small-pool MIRACL/zh case. The same pattern holds across English (TREC, AOL, ORCAS) and Chinese (MIRACL/zh), so the \systemname{} B2 protocol works across both data types and languages.

\begin{table}[t]
\centering
\small
\caption{B2 tester agreement (Kendall's $\tau$ vs.\ qrels-nDCG@10) across four datasets in two languages. 10 retrieval systems; 95\% CI from 1000 bootstrap resamples.}
\label{tab:b2}
\begin{tabular*}{\columnwidth}{@{\extracolsep{\fill}}lcccc@{}}
\toprule
Tester (sim.) & TREC S'14 & AOL-IA & ORCAS & MIRACL/zh \\
\midrule
SimIIR-PBM & 0.778 & 0.822 & 0.911 & 0.911 \\
SimIIR-DBN & 0.822 & 1.000 & 1.000 & 0.867 \\
Heuristic  & $-0.067$ & 0.067 & $-0.111$ & 0.200 \\
LLM-sim    & 0.956 & 0.956 & 0.956 & 0.867 \\
\bottomrule
\end{tabular*}
\vspace{-0.6em}
\end{table}

On AOL-IA, RATE-style aggregation gives reliability weights of 1.208 (SimIIR-DBN), 1.204 (LLM-sim), 1.204 (qrels reference), 1.199 (SimIIR-PBM), and 0.185 (Heuristic). Leave-one-out sensitivity keeps BERT\_CAT as the top system in every configuration on every dataset, so the top-of-list conclusion is robust on this set of retrieval systems. The middle-ranked systems still move noticeably when individual simulators are dropped. The Chinese MIRACL/zh result shows a second effect: the small judged-pool case (8--15 docs/query) lifts the Heuristic baseline's $\tau$ to 0.20 because a position-1 click is fairly likely to land on a relevant doc when only 1--2 of the 8--15 candidates are relevant.

\subsection{Realism Does Not Equal Reliability (B3)}

B3 measures how well B1 realism metrics predict B2 tester agreement across $n{=}48$ simulator-seed pairs (four simulators on the two retrieval-grounded session datasets, six seeds each). Table~\ref{tab:b3} reports pooled and per-dataset Pearson correlations.

\begin{table}[t]
\centering
\small
\caption{B3: correlation between realism metrics (B1) and tester agreement (B2). Pooled aggregates TREC Session'14 and AOL-IA (n=48); AOL is the AOL-IA shard alone (n=24). Bold marks significant correlations ($p{<}0.05$); negative $r$ for distance metrics is expected.}
\label{tab:b3}
\begin{tabular*}{\columnwidth}{@{\extracolsep{\fill}}lcccc@{}}
\toprule
Realism metric (B1) & Pooled $r$ & $p$ & AOL $r$ & $p$ \\
\midrule
JS (click depth)                          & $\mathbf{-0.43}$ & 0.002 & $\mathbf{-0.75}$ & ${<}10^{-4}$ \\
Fr\'{e}chet (embed) \cite{Heusel2017FID}  & $\mathbf{-0.40}$ & 0.005 & $\mathbf{-0.75}$ & ${<}10^{-4}$ \\
Reform.\ similarity                       & $\mathbf{+0.36}$ & 0.012 & $+0.38$          & 0.070 \\
JS (action types)                         & $\mathbf{-0.31}$ & 0.032 & $\mathbf{-0.69}$ & ${<}10^{-3}$ \\
MMD (embed) \cite{Gretton2012MMD}         & $\mathbf{-0.31}$ & 0.033 & $\mathbf{-0.64}$ & ${<}10^{-3}$ \\
W (session length)                        & $-0.22$          & 0.136 & $\mathbf{-0.52}$ & 0.010 \\
Classifier (audited)                      & $+0.09$          & 0.563 & $\mathbf{+0.51}$ & 0.010 \\
\bottomrule
\end{tabular*}
\vspace{-0.6em}
\end{table}

\begin{figure}[t]
    \centering
    \includegraphics[width=\linewidth]{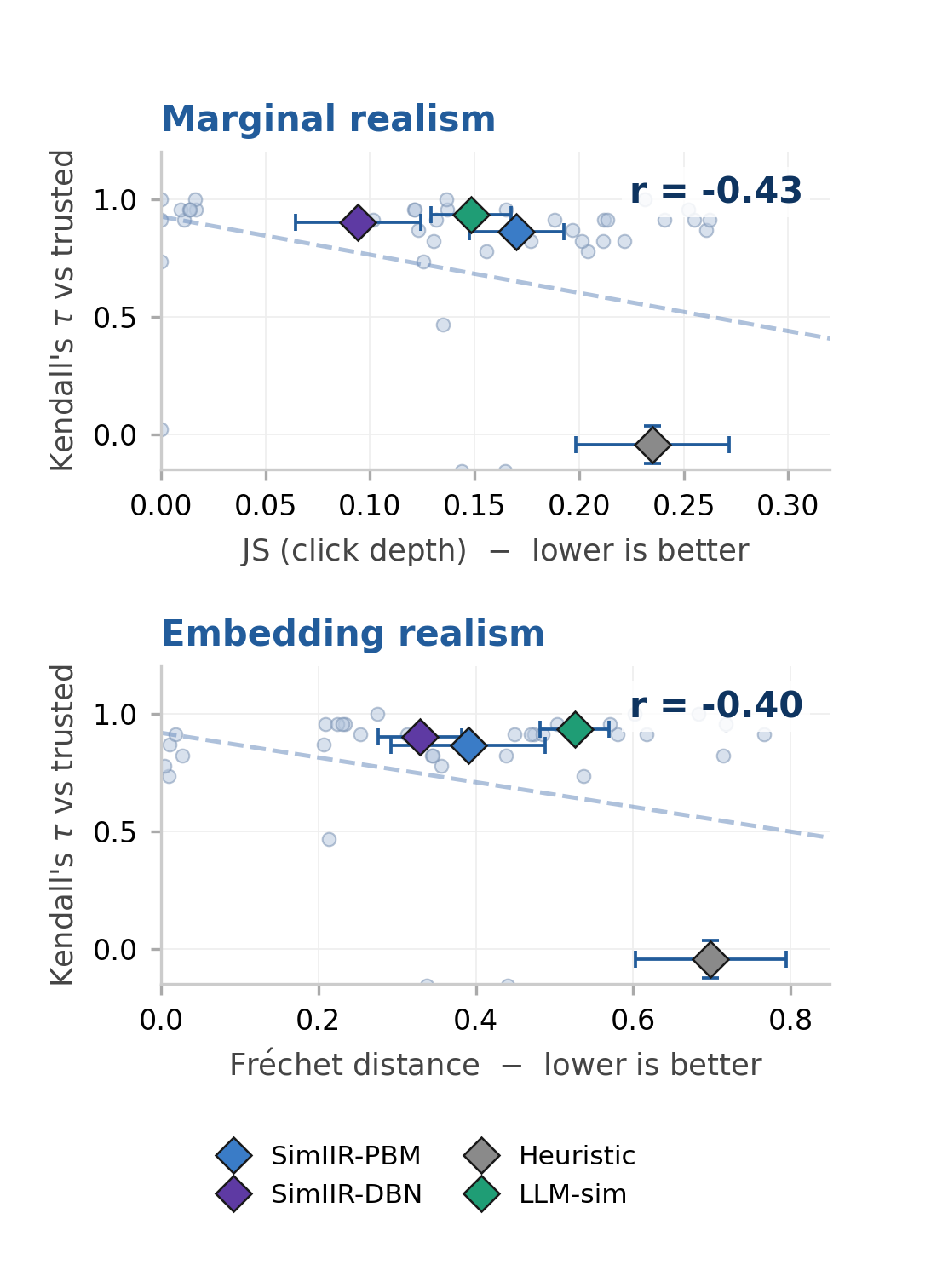}
    \caption{Realism does not equal reliability. Each panel plots Kendall's $\tau$ (system-ranking validity) against one B1 realism distance. Diamonds are per-simulator means; the corner $r$ is the Pearson correlation across all 48 (simulator, dataset, seed) points. The Heuristic sits far below the trend in both panels: similar realism distance to the relevance-aware simulators, but $\tau$ near zero.}
    \vspace{-1.6em}
    \label{fig:tradeoff}
\end{figure}

Three findings stand out. First, click-depth distance is the single strongest predictor across pooled and per-dataset analyses ($|r|{=}0.43$ pooled, 0.75 on AOL alone). Second, Fr\'{e}chet distance over session embeddings is a close second ($|r|{=}0.40$ pooled, 0.75 on AOL alone), and it captures click-position structure and higher-order action-transition patterns. Third, the per-dataset gap is large: TREC Session'14 alone ($n{=}24$) gives no significant correlation for any distance metric, because each per-seed shard has only $\sim$200 sessions and the B1 estimates are noisy. Pooling lifts five of seven realism metrics to significance. The audited classifier has a strong AOL-only signal ($r{=}+0.51$) but the pooled effect collapses to $r{=}+0.09$, which means the classifier picks up dataset-specific stylistic quirks and not a transferable behavioural difference. Reformulation similarity is positive in both columns but its per-dataset $p{=}0.07$ keeps it marginal. Figure~\ref{fig:tradeoff} plots the two strongest predictors and shows the simulators in different regions of the realism/reliability plot.

\textbf{The validity question, made measurable.}
Calling a simulator ``human-like'' is a weaker claim than showing that the system rankings it produces match those of real users: a simulator can pass every behavioural check and still favour particular system designs. Table~\ref{tab:b3} puts numbers on this gap. The strongest realism predictor (JS click depth) accounts for at most $\sim$56\% of the variance in tester agreement on AOL-IA ($r^2{=}0.56$) and $\sim$18\% pooled across datasets. The classifier-discriminator score, the dominant ``human-likeness'' check in the simulator literature, collapses to $r{=}+0.09$ ($p{=}0.56$) pooled, no better than a coin flip; its AOL-only signal at $r{=}+0.51$ does not transfer. Click-depth distance is roughly five times more predictive of validity than the classifier score, yet is rarely reported. The Heuristic baseline that ignores relevance produces $\tau<0.21$ on every corpus while still emitting plausible click traces, a concrete case where a plausible-looking simulator gets system rankings wrong. As a worked example: a paper that reports an AOL-IA classifier AUC near 0.91 and stops there gives no evidence that its simulator is a valid tester. SimIIR-PBM and the Heuristic both hit AUC near 0.91 in Table~\ref{tab:b1} (0.910 and 0.919); against the same qrels in Table~\ref{tab:b2}, PBM scores $\tau{=}0.82$ (valid as a tester) and the Heuristic scores $\tau{=}0.07$ (useless).

\textbf{Practical implications.}
Realism is necessary but not enough for tester validity, and which realism metric matters depends on the corpus. JS divergence on click depth is informative on web log data but breaks down on conversational corpora where ``click depth'' is undefined. Fr\'{e}chet distance \cite{Heusel2017FID} and MMD \cite{Gretton2012MMD} give a consistent signal across data types, at the cost of needing an embedder. Reformulation-based scores should not be used on logs where verbatim reformulation is rare. Researchers using simulators as testers should measure ranking agreement against a trusted evaluator directly when they can: even our best behavioural simulator was an unreliable tester on the smallest-sample case ($n{=}10$ systems, 100 queries). For practitioners working without TripClick or TianGong-ST access, the cross-corpus B2 results in Table~\ref{tab:b2} support ORCAS \cite{Craswell2020ORCAS} (English) and MIRACL/zh \cite{Zhang2023MIRACL} (Chinese) as drop-in alternatives.

\section{Access and Reproducibility}
\label{sec:access}

\systemname{} is released as an open-source Python package under MIT license. The release includes: (R1)~the core library with schema, adapters, metrics, and benchmark runners; (R2)~configuration files for each official benchmark (e.g.\ \texttt{B1-AOL-IA-v1}, \texttt{B2-AOL-IA-v1}) that can be cited in future work; (R3)~the protocol driver that reproduces every number in Section~\ref{sec:results} given the listed seeds; and (R4)~documentation with adapter specifications, simulator integration recipes, and interpretation guidance. \systemname{} does not redistribute raw datasets; adapters and download helpers are provided for users to obtain data from original sources under their licences. Every benchmark run emits provenance (dataset version, configuration hash, schema version, random seeds), so any reported number can be traced to its configuration and regenerated.

\textbf{What \systemname{} decides for you.}
Three choices are usually buried in custom evaluation scripts and rarely line up across papers: (i)~the sessionisation threshold and how it interacts with click-depth metrics; (ii)~which classifier features are admissible (and which are pure adapter signature); (iii)~how the trusted tester is constructed for B2 (positive-only qrels with distractor padding versus full-pool qrels). \systemname{} commits to one default for each, records the choice in the manifest, and exposes the alternative as a configuration override. A simulator paper that cites a configuration ID (e.g., \texttt{B2-AOL-IA-v1}) thereby commits to a precise comparison surface other papers can replicate.

\section{Ethics and Limitations}
\label{sec:ethics}

Interaction logs can contain sensitive information. \systemname{} does not introduce new data collection but it does lower the barrier to analysis. The safeguards are: no cross-dataset user linkage, optional text sanitisation, and aggregate-only default outputs. Interaction data carries the biases of the populations and ranking policies that produced it, and simulators tuned to agree with biased testers may amplify those biases.

Several limitations bound the framework itself. The canonical schema is built from discrete events: continuous signals like eye-tracking, hover, and mouse movement fall outside its scope, and dialogue acts beyond the listed types need schema extensions before B1 can score them. The simulator pool included in the baselines covers click models (PBM, DBN), a position-only heuristic, and an LLM-style click sampler; query-reformulation simulators built from contextual Markov models of query change \cite{Zerhoudi2021QueryChange}, and tool-using LLM agents that interleave retrieval calls with their reasoning, are open additions on top of the same adapter interface. Realism metrics depend on the embedder; the default action-sequence embedder is one option among many the protocol supports \cite{Reimers2019SBERT}, and we have not compared them systematically. Tester reliability is measured against qrels-nDCG@10, which is itself a proxy that shifts as pools are updated and judgements age \cite{Parry2025Variations}; \systemname{} cannot escape that ceiling. Within these bounds, B3 ($n{=}48$) has enough samples to detect pooled effects, but per-seed shards smaller than $\sim$200 sessions remain too noisy for stable per-dataset B1 estimates.

By standardising simulator evaluation, \systemname{} may indirectly speed up the development of simulators whose synthetic traces are hard to tell from real ones, with potential for misuse in click fraud or search manipulation. \systemname{} is built for evaluation, not generation; the schema-baseline and permutation comparisons that bound classifier AUC double as a diagnostic for detecting synthetic traffic in production, which works in the other direction.

\section{Conclusion}
\label{sec:conclusion}

We presented \systemname{}, an open-source toolkit and benchmark suite for simulator-centric evaluation in interactive IR. \systemname{} standardises session representation, dataset preprocessing, and evaluation protocols under two distinct objectives, behavioural realism and tester reliability, with validity rules checked at run time and full provenance for every result. Baseline numbers on four real datasets across two languages (TREC Session Track 2014, AOL-IA, ORCAS, MIRACL/zh) show that behavioural realism and tester reliability are correlated but not interchangeable: click-depth distance ($|r|{=}0.43$ pooled, 0.75 on AOL-IA alone) and Fr\'{e}chet distance over session embeddings ($|r|{=}0.40$ pooled, 0.75 on AOL-IA alone) are the strongest predictors of ranking agreement ($p{\leq}0.005$, $n{=}48$), while the audited classifier transfers poorly. The same B2 protocol works for English and Chinese without modification.

Three follow-on questions sit on top of this foundation. First, do realism metrics that explicitly model session-internal dependencies (Markov-chain agreement on action sequences, inter-event timing) close the pooled-classifier gap? Second, does an LLM agent with retrieval and tool use rank systems any more reliably than the click-model simulators, or do its surface differences from real users (verbose queries, deeper examination) cancel its semantic advantage? Third, how much qrels coverage is enough: at what number of judged docs per query does the B2 trusted tester stop biasing the ranking it should be checking? \systemname{} is the lowest-friction way to run any of these experiments; we invite the community to extend it with new datasets, metrics, and simulator adapters.





\bibliographystyle{ACM-Reference-Format}
\bibliography{sample-base}

\end{document}